\documentclass[12pt]{iopart}

\usepackage{iopams}
\usepackage{amssymb,amsfonts}
\usepackage[dvips]{graphicx}
\usepackage{psfrag}

\begin{document}
\bibliographystyle{unsrt}

\title[Non-Hermitian Quantum  Annealing  \dots]{
Non-Hermitian Quantum  Annealing in the Antiferromagnetic Ising Chain}

\author{Alexander I. Nesterov }

\address{Departamento de F{\'\i}sica, CUCEI, Universidad de Guadalajara,
Av. Revoluci\'on 1500, Guadalajara, CP 44420, Jalisco, M\'exico}


\author{Gennady P.  Berman}

 \address{Theoretical Division, Los Alamos National Laboratory,
Los Alamos, NM 87544, USA }
\ead{gpb@lanl.gov}

\author{Juan C. Beas Zepeda}

\address{Departamento de F{\'\i}sica, CUCEI, Universidad de Guadalajara,
Av. Revoluci\'on 1500, Guadalajara, CP 44420, Jalisco, M\'exico}
 \ead{juancarlosbeas@gmail.com}

 \author{Alan R.~Bishop}

\address{Los Alamos National Laboratory, STE, MS A127, Los Alamos,  NM, 87544, USA}
 \ead{ arb@lanl.gov}

\begin{abstract}

 A non-Hermitian quantum optimization algorithm is created and used to find the ground state of an antiferromagnetic Ising chain.  We demonstrate analytically and numerically (for up to $N=1024$ spins) that our approach leads to a significant reduction of the annealing time that is proportional to $\ln N$, which is much less than the time (proportional to $N^2$) required for the  quantum annealing based on the corresponding Hermitian algorithm. We propose to use this approach to achieve similar speed-up for NP-complete problems by using classical computers in combination with quantum algorithms.

 ~~~~~~~~~~~~~~~~\\
 {\em  LA-UR: 13-21324}\\

\end{abstract}

\submitto{\NJP}



\maketitle

\tableofcontents


\section{Introduction}

It is generally recognized that quantum annealing (QA) algorithms can find applications for solving many practically important problems including optimization of complex networks, finding the global minimum of multi-valued functions, and cost minimization. Instead of classical annealing algorithm in which the temperature effects are utilized, the QA can operate, in principle, at zero temperature. In this case, one first formulates the optimization problem in terms of finding the ground state of the effective quantum Hermitian Hamiltonian, $\mathcal H_0$, which describes a quantum time-evolution of a quantum register with $N$ qubits. During the slow and directed time-evolution, the quantum system tunnels to the global ground state of this effective Hamiltonian. One of the main challenges is to accelerate the speed of QA algorithms, so that the annealing time grows not exponentially, but polynomially with the size of the quantum register \cite{KN,FGGLL,SSMO,DC,SMTC,SST,SNS,AM}.

Many approaches dealing with finding the ground state of the Hamiltonian, $\mathcal H_0$, using QA algorithms, can be found in \cite{KN,FGGLL,SSMO,DC,SMTC} (see also references therein). To illustrate the idea of QA algorithms, consider the time-dependent Hamiltonian,
$\mathcal H(t) =\mathcal H_0 +  \Gamma(t)\mathcal H_1$, where $\mathcal H_0$ is the Hamiltonian to be optimized and $\mathcal H_1$ is an auxiliary (``initial") Hamiltonian. In order to initiate a non-trivial quantum dynamics of a register, the condition, $[\mathcal H_0,\mathcal H_1] \neq 0$, is imposed.  The external time-dependent field, $\Gamma(t)$, decreases from a large enough value to zero during the quantum annealing. The ground state of $\mathcal H_1$ is the initial state, which is assumed to be known. Then, the adiabatic theorem guarantees that if $\Gamma(t)$ decreases sufficiently slow,  the system approaches the ground state of $\mathcal H_0$, at the end of the quantum annealing.

One of the requirements of the adiabatic theorem is the existence of a finite gap between instantaneous ground state and first excited state of the total Hamiltonian, $\mathcal H(t)$. However, usually in quantum optimization of $N$-qubit models the minimal energy gap becomes exponentially small, producing an exponential increase of the annealing time \cite{FGGLL,DC,SUD,JKKM,YKS}. Then, the main problem is on how to speed up the performance of the QA algorithms and, at the same time, find the ground state of $\mathcal H_0$.

In  \cite{BN}, we proposed an adiabatic quantum optimization algorithm based on non-Hermitian quantum mechanics. Our approach demonstrated that when using the non-Hermitian term, $\mathcal H_1$, an effective level repulsion takes place for the total Hamiltonian, greatly increasing the minimum energy gap.

In \cite{NABG,NABJBG}, we applied this non-Hermitian quantum annealing (NQA) algorithm to Grover's problem of finding a marked item in an unsorted database, and to study the transition to the ground state in a 1-dimensional ferromagnetic Ising chain. We demonstrated that, for a suitable choice of relaxation and other parameters, the search time is proportional to the logarithm of the number of qubits (spins).

In this paper, we apply our previously developed NQA to study the transitional dynamics to the ground state of $\mathcal H_0$ in a 1-dimensional antiferromagnetic Ising spin chain in a time-dependent transverse magnetic field. We also assume that the dissipation vanishes at the end of evolution. When the annealing is completed, the system is governed by the Hermitian Hamiltonian, $\mathcal H_0$.

The main reason for choosing the antiferromagnetic Ising spin chain is that the ground state in this model is much more complicated than in the corresponding ferromagnetic chain. Namely, it includes rapid spatial oscillations (correlations) of spins. This allows us to demonstrate that our (NQA) algorithm works well for this case, and significantly decreases  the annealing time. Another reason is to consider the behavior of quantum correlation functions (which are more complicated than in the ferromagnetic chain) and  formation of defects generated in the process of annealing.

We show that the NQA significantly increases the probability for the system to remain in the ground state. In particular, a comparison with the results of the Hermitian QA reveals that the NQA reaches the ground state of $\mathcal H_0$ with much larger probability, when we use the same annealing scheme for both. We show that the NQA has a complexity of order $\ln N$, where $N$ is the number of spins. This is much better than the quantum Hermitian adiabatic algorithm which has the complexity for this problem of order $N^2$.

This paper is organized as follows. In Sec. II, we introduce a dissipative one-dimensional antiferromagnetic Ising system in a transverse magnetic field governed by a non-Hermitian Hamiltonian. In Sec. III, we study the quench dynamics of our system using both analytic and numerical methods. In Sec. IV, we study the correlation properties of our system and defects formation at the end of evolution.  We conclude in Sec. V with a discussion of our results. In the Appendix we present some technical details.

\section{Description of the model}

We consider the 1-dimensional antiferromagnetic Ising chain in a transverse magnetic field with dissipation governed by the following non-Hermitian Hamiltonian:
\begin{eqnarray} \label{EqH1}
H=  \frac{J}{2}\sum^N_{n=1}\big(g \sigma^x_n + \sigma^z_n \sigma^z_{n+1} - i2{\delta}\sigma^{+}_n \sigma^{-}_n \big),
\end{eqnarray}
with the periodic boundary condition, $\mbox{\boldmath$\sigma$}_{N+1} =\mbox{\boldmath$\sigma$}_1$. The external magnetic field is associated with the parameter, $g$, the spontaneous decay is described by the parameter,  $\delta$, and $\sigma^{\pm}_n =  (\sigma^{z}_n \pm i \sigma^{y}_n)/2$ are the spin raising and lowering operators.

In principle, this model can be realized in a chain of the superconducting phase qubits with the tunneling to the continuum mostly from the lowest energy levels. We apply the standard Feshbach projection method to obtain the effective non-Hermitian Hamiltonian. The motivations for choosing the non-Hermitian relaxation term in Eq. (\ref{EqH1}) are discussed in the Conclusion.

The Hamiltonian in Eq. (\ref{EqH1}) can be diagonalized using the standard Jordan-Wigner transformation, following procedures outlined in \cite{NABJBG,KSH,DJ,CP,NO}. The Jordan-Wigner transformation, which is mapping of a spin-1/2 system to a system of spinless fermions, is given by
\begin{eqnarray}\label{Eq1c}
\sigma^x_n = 1-2 c^\dagger_n c_n, \\
\sigma^y_n = i(c^\dagger_n - c_n  )\prod_{m<n}(1-2 c^\dagger_m c_m), \\
\sigma^z_n = -(c_n + c^\dagger_n )\prod_{m<n}(1-2 c^\dagger_m c_m),
\end{eqnarray}
in which the $c_n$ are fermionic operators that satisfy anticommutation relations:
 $\{c^\dagger_m,c_n\} = \delta_{mn}$ and $\{c_m,c_n\}=\{c^\dagger_m,c^\dagger_n\}=0$.
Applying these transformations, we obtain
\begin{eqnarray}\label{I}
\fl
H = \displaystyle\frac{J}{2}\sum^N_{n=1}\big(c^\dagger_n c_{n+1 }+ c^\dagger_{n+1}c_{n}+ c_{n +1}c_{n }+ c^\dagger_{n} c^\dagger_{n+1}+ g(1-2 c^\dagger_n c_{n }) - 2i\delta c^\dagger_n c_{n }  \big).
\end{eqnarray}

The periodic boundary condition imposed on the spin operators leads to the following boundary condition for the fermionic operators:
 \begin{eqnarray}\label{Eq1b}
  c_{N+1} = - e^{i\pi {\mathcal N}_F} c_1,
 \end{eqnarray}
${\mathcal N}_F = \sum^N_{n=1}  c^\dagger_{n} c^\dagger_{n}$ being the total number of fermions. From Eq. (\ref{Eq1c}) it follows that ${\mathcal N}_F=  N/2- S^x$, where $S^x$ is the total $x$-component of the spins. For the particular choice of parameters, $g=\delta=0$, we obtain, $S^x =0$ and ${\mathcal N}_F=  N/2$. Thus, for $S^x =0$ and ${\mathcal N}_F=  N/2$ we obtain {\em periodic} ({\em antiperiodic}) boundary conditions if $N/2$ is {\em odd} ({\em even}). Note that since the parity of the fermions is conserved, the imposed boundary conditions are valid for all values of the parameters, $g$ and $\delta$.

Next, by applying the Fourier transformations,
\begin{eqnarray}
c_n = \frac{e^{-i\pi/4}}{\sqrt{N}}\sum_k c_k e^{i2\pi kn/N},
\end{eqnarray}
we find that the Hamiltonian (\ref{I}) can be recast in Fourier space as follows:
\begin{eqnarray}\label{Ia}
H= -\sum_{k}\frac{J}{2}\Big(2(\tilde g - \cos \varphi_k)c^\dagger_k c_k-  g
+ \sin \varphi_k(c^\dagger_k c^\dagger_{-k} + c_{-k} c_{k} ) \Big),
\end{eqnarray}
where $\tilde g = g+i\delta$ and $\varphi_k = {2\pi k}/{N}$. For periodic boundary conditions, $c_{N+1}=c_1$, the wave number, $k$, takes the following discrete values:
  \begin{eqnarray}\label{Eq1ar}
 k=-\frac{N}{2}, \dots, 0,1, \dots,  \frac{N}{2}-1,
 \end{eqnarray}
and for antiperiodic boundary conditions, $c_{N+1}=-c_1$, we obtain
  \begin{eqnarray}\label{Eq1a}
 k=\pm\frac{1}{2},\pm\frac{3}{2},\dots, \pm\frac{N-1}{2},
  \end{eqnarray}
In what follows, we impose the
antiperiodic boundary conditions for the fermionic operators.

The Hamiltonian, $H$, can be diagonalized using the generalized Bogoliubov transformation \cite{NO}. Its spectrum is given by $\varepsilon_{\pm}(k)= \varepsilon_{0} \pm \varepsilon_k$, in which $\varepsilon_{0} = J\cos \varphi_k- iJ\delta $, and $ \varepsilon_k = J\sqrt{\tilde g^2 -  2\tilde g\cos \varphi_k +1}$. There are two (right) eigenstates for each $k$,
\begin{eqnarray}\label{E2a}
 &|u_{+}(k)\rangle = \left(\begin{array}{c}
                 \cos\frac{\theta_k}{2} \\
                  \sin\frac{\theta_k}{2}
                  \end{array}\right), \\
&|u_{-}(k)\rangle = \left(\begin{array}{c}
               -\sin\frac{\theta_k}{2}\\
                 \cos\frac{\theta_k}{2}
                  \end{array}\right),
                  \label{E2b}
 \end{eqnarray}
in which
\begin{eqnarray}\label{Th}
\cos\theta_k = \frac{  \cos \varphi_k - \tilde g}{\sqrt{\tilde g^2 - 2\tilde g\cos \varphi_k +1}}, \\
\sin\theta_k =- \frac{ \sin \varphi_k}{\sqrt{\tilde g^2 -  2\tilde g\cos \varphi_k +1}},
\end{eqnarray}
with $\theta_k$ being a complex angle.

Since for each $k$, the ground state lies into the two-dimensional Hilbert space spanned by $|0\rangle_k  |0\rangle_{-k}$ and $|1\rangle_k  |1\rangle_{-k}$, it is sufficient to project $H_k$ onto this subspace. For a given value of $k$, both of these states can be represented as a point on the complex two-dimensional sphere, $S^2_c$. In this subspace, the Hamiltonian, $H_k$, takes the form
\begin{eqnarray}\label{H1a}
 {H}_k = {\varepsilon_0} {1\hspace{-.125cm}1}- {J} \left(
\begin{array}{cc}
             \tilde g- \cos \varphi_k & \sin \varphi_k \\
             \sin \varphi_k & -\tilde g+ \cos \varphi_k  \\
            \end{array}
          \right).
\end{eqnarray}
Its ground state can be written as a product of qubit-like states, $|\psi_g\rangle =  \bigotimes_{k}|u_{-}(k)\rangle $, so that:
\begin{eqnarray}
&|\psi_g\rangle =  \bigotimes_{k} \Big(\cos\frac{\theta_k}{2}|0\rangle_k  |0\rangle_{-k}  -\sin\frac{\theta_k}{2}|1\rangle_k  |1\rangle_{-k} \Big),
\end{eqnarray}
where, $|0\rangle_k $, is the vacuum state of the mode $c_k$, and $|1\rangle_k  $ is the excited state: $|1\rangle_k =c^\dagger_k |0\rangle_k$.

For $|\tilde g|\gg 1$, the ground state is paramagnetic with all spins oriented along the $x$ axis. From Eq. (\ref{Th}) we obtain $\cos\theta_k \rightarrow- 1$ as $|\tilde g| \rightarrow \infty$. From here it follows that, $|u_{-}(k)\rangle \rightarrow \scriptsize \left(\begin{array}{c}
 1 \\
  0 \\
  \end{array}
  \right)$.
Thus, the north pole of the complex Bloch sphere corresponds to the paramagnetic ground state. On the other hand, when $|g|\ll 1$ there are two degenerate antiferromagnetic ground states with neighboring spins polarized in opposite directions along the $z$-axis. The real part of the complex energy reaches its minimum at the point defined by $\cos\theta_k = 1$, and, hence, the south pole of the complex sphere is related to the pure antiferromagnetic ground state.

\section{ Quantum annealing}

In this section, we consider the NQA for the time-dependent Hamiltonian of Eq. (\ref{EqH1}) written as
\begin{eqnarray}\label{EqH2}
 \tilde{\mathcal H}_\tau(t)= {\mathcal H}_0 + {\mathcal H}_1(t),
\end{eqnarray}
where
\begin{eqnarray}
&{\mathcal H}_0 =  \frac{J}{2}\sum^N_{n=1}\sigma^z_n \sigma^z_{n+1}, \\
&{\mathcal H}_1(t) = \frac{J}{2}\sum^N_{n=1}\big(g(t) \sigma^x_n - i2\delta(t)\sigma^{+}_n\sigma^{-}_n \big).
\end{eqnarray}

We use with the ground state of the auxiliary Hamiltonian, ${\mathcal H}_1(0)$, as the initial state. This state is ``paramagnetic'' with all spins oriented along the $x$ axis. For $g \gg 1$ the Hamiltonian, $\mathcal H_\tau(0)$, is dominated by $\mathcal H_1(0)$, and the ground state of the total Hamiltonian, $ \tilde{\mathcal H}_\tau$, is determined by the ground state of  $\mathcal H_1(0)$. The term, ${\mathcal H}_1$, causes quantum tunneling between the eigenstates of the Hamiltonian, ${\mathcal H}_0$. At the end of the NQA, we obtain, $ \tilde{\mathcal H}_\tau(\tau)= {\mathcal H}_0 $.  If the quench is sufficiently slow, the adiabatic theorem guarantees reaching the ground state of the main Hamiltonian, $\mathcal H_0$, at the end of computation.

As shown in Sec. II, the total Hamiltonian, $ \tilde{\mathcal H}_\tau(t)$, in the momentum representation splits into a sum of independent terms, $  \tilde{\mathcal H}_\tau(t)= \sum_{k}\mathcal H_k(t)$. Each $\mathcal H_k$ acts in the two-dimensional Hilbert space spanned by $|k_1\rangle =  |1\rangle_k  |1\rangle_{-k}$ and $ |k_0\rangle= |0\rangle_k  |0\rangle_{-k}$. The wavefunction can be written as, $|\psi(t)\rangle = \prod_{k}|\psi_k(t)\rangle$, where
\begin{eqnarray}\label{Eq10}
|\psi_k(t)\rangle=   c_0(k,t)|k_0\rangle + c_1(k,t) |k_1\rangle.
\end{eqnarray}

Choosing the basis as, $k_1= \scriptsize\left(
                              \begin{array}{c}
                                1 \\
                                0 \\
                              \end{array}
                            \right)$
and $k_0=\scriptsize\left( \begin{array}{c}
                                0 \\
                                1 \\
                              \end{array}
                            \right)$,
we find that the Hamiltonian, $ \mathcal {H}_k(t)$, projected on this two-dimensional subspace takes the form
\begin{eqnarray}\label{H1g}
 \mathcal {H}_k(t)  = {\varepsilon_0(t)} {1\hspace{-.125cm}1} - {J} \left(
\begin{array}{cc}
              \tilde g(t)- \cos \varphi_k & \sin \varphi_k \\
             \sin \varphi_k & -\tilde g(t)+ \cos \varphi_k  \\
            \end{array}
          \right),
\end{eqnarray}
where $\varepsilon_0(t)= J\cos \varphi_k- iJ\delta(t)$ and $\tilde g(t) = g(t) + i\delta(t)$. Further we assume linear dependence of $\tilde g(t)$ on time:
\begin{eqnarray}
\tilde g(t) = \left\{
\begin{array}{l}
\gamma (\tau-t), \quad 0 \leq t \leq \tau \\
0 , \quad t > \tau
\end{array}
\right.
\end{eqnarray}
where $\gamma=(g +i\delta)/\tau$, and $g$, $\delta$ are real parameters.

The general wave functions,  $|\psi_k\rangle$ and  $\langle\tilde \psi_k |$, satisfy the Schr\"odinger equation and its adjoint equation
 \begin{eqnarray}\label{Eqh1}
i\frac{\partial }{\partial t}|\psi_k\rangle&= \mathcal {H}_k(t)|\psi_k\rangle , \\
-i\frac{\partial }{\partial t}\langle\tilde \psi_k |& =
\langle\tilde \psi_k |\mathcal {H}_k(t)\label{NS2}.
\end{eqnarray}
Presenting $|\psi_k(t)\rangle$ as a linear superposition,
\begin{eqnarray}\label{S1}
|\psi_k(t)\rangle = (u_k(t)|k_0\rangle + v_k(t)|k_1\rangle) e^{i\int \varepsilon_0(t)dt},
\end{eqnarray}
and inserting expression (\ref{S1}) into Eq. (\ref{Eqh1}), we obtain
\begin{eqnarray}\label{IS2a}
i\dot  u_k &= J\big((\tilde  g - \cos \varphi_k)\,u_k - \sin \varphi_k\, v_k\big), \\
i\dot  v_k &= -J\big(\sin \varphi_k\, u_k +(\tilde  g - \cos \varphi_k)\,v_k \big ).
\label{IS2b}
\end{eqnarray}
The solution can be written in terms of the parabolic cylinder functions, $D_{\nu}(z)$:
\begin{eqnarray}\label{Aq4}
& U_{k}(z_k)= B_k D_{i\nu_k}(iz_k) +{\sqrt{i\nu_k }}  A_k D_{-i\nu_k-1}(z_k)  ,\\
& V_{k}(z_k)=   A_k D_{-i\nu_k}(z_k) -i\sqrt{i\nu_k } B_k D_{i\nu_k-1}(iz_k))  .
\label{Aq4a}
 \end{eqnarray}
Here we introduced the new functions: $u_k(t)= U(z_k)$, $v_k(t) = V(z_k)$ and
\begin{eqnarray}\label{Eq4b}
&z_k(t) = e^{i\pi/4}\sqrt{\frac{2J}{\gamma}}\Big(\gamma(\tau -t)-\cos \varphi_k\Big), \\
&\nu_k= \frac{J\sin^2 \varphi_k}{2\gamma}.
\label{Eq4c}
\end{eqnarray}
The constants, $A_k$ and $B_k$, in Eqs. (\ref{Aq4}) and (\ref{Aq4a}) are determined by the initial conditions.

In the adiabatic basis the wavefunction, $|\psi_k(t) \rangle$, can be written as follows:
\begin{eqnarray}\label{Eq7}
 |\psi_k(t) \rangle =( \alpha_k(t) |u_{-}(k,t)\rangle + \beta_k(t) |u_{+}(k,t)\rangle)e^{i\int \varepsilon_0(t)dt}.
\end{eqnarray}
From Eqs. (\ref{S1}) and (\ref{Eq7}) it follows that
\begin{eqnarray}\label{Eq6a}
\alpha_k(t) =u_k(t)\cos\frac{\theta_k(t)}{2}- v_k(t)\sin\frac{\theta_k(t)}{2},
\end{eqnarray}
and
\begin{eqnarray}\label{Eq6ar}
\beta_k(t) =v_k(t)\cos\frac{\theta_k(t)}{2}+ u_k(t)\sin\frac{\theta_k(t)}{2}.
\end{eqnarray}

We assume that the evolution begins from the ground state. Then, for any $k$ this implies, $|\alpha_k(0)|=1$ and $\beta_k(0)=0$. Since $\cos\theta_k(0)=-1$, one has, $u_k(0)=0$ and $v_k(0)=-1$. Using these results, we obtain $B_k=0$, and the solution of the Schr\"odinger equation can be written as follows:
\begin{eqnarray}\label{eq5a}
& U_{k}=  A_k {\sqrt{i\nu_k }} D_{-i\nu_k-1}(z_k),\\
& V_{k}=  A_k D_{-i\nu_k}(z_k),
\label{eq5b}
 \end{eqnarray}
in which $A_k= (D_{-i\nu_k}\big(z_k(0)\big))^{-1}$.

Since for non-Hermitian systems the norm of the wavefunction is not conserved, for each $k$ we define the ``intrinsic" probability to stay in the ground state as
\begin{eqnarray}\label{QEq2}
 P^{gs}_k(t)= \frac{|\alpha_k(t)|^2}{|\alpha_k(t)|^2 + |\beta_k(t)|^2}.
\end{eqnarray}
For each $k$ the evolution of the system is independent, and the probability for the whole system to stay in the ground state is given by $P_{gs}(t)= \prod_k P^{gs}_k(t)$.
In what follows, we use the asymptotic expansion for Weber functions to calculate the probability, $P_{gs}= \prod_k P^{gs}_k(\tau)$, to remain in the ground state at the end of evolution,

For long wavelength modes with $|\varphi_k|\ll \pi/4$,  using Eqs. (\ref{eq5a}), (\ref{eq5b}), we obtain
\begin{eqnarray}\label{Eq2a}
 P^{gs}_k(\tau)\approx \frac{1}{1+ \displaystyle\frac{| D_{-i\nu_k}(z_k(\tau)) |^2}{|\sqrt{i\nu_k}\, D_{-i\nu_k-1}(z_k(\tau)) |^2}}.
\end{eqnarray}
Using the asymptotic formulas for Weber functions with the large argument values, and taking into account that $\delta\ll g$, we obtain, as in \cite{NABJBG},
\begin{eqnarray}\label{Pq2k}
P_k(\tau) =\frac{1-  e^{-2\pi\Re\nu_k}}{1-e^{-2\pi\Re\nu_k}+  e^{-2\pi\Re\nu_k- \Re z^2_k(\tau)}},
\end{eqnarray}
where
\begin{eqnarray}\label{Eq3a}
 &\Re\nu_k = J\tau \pi^2 k^2 /(2gN^2) ,  \\
& \Re z^2_k(\tau) = 2\delta J\tau/g^2.
 \label{Eq3b}
\end{eqnarray}
For the Hermitian QA, $\delta =0$, and Eq. (\ref{Pq2k}) leads to the Landau-Zener formula \cite{LL,ZC},
\begin{eqnarray}\label{P1}
P_k(\tau) =1-\exp\bigg(-\frac{J\tau \pi^3 k^2}{g N^2 }\bigg).
\end{eqnarray}

Similar considerations of the short wavelength modes with $3\pi/4 <|\varphi_k|\leq \pi$ yields, $P^{gs}_k(\tau) = 1 + {\mathcal O}(\sqrt{|\nu_k|})$. For short wavelength modes,  with $\pi/4 < |\varphi_k| \leq \pi/2$, employing the large-order asymptotics for Weber functions, we obtain, $P^{gs}_k(\tau) = 1 + {\mathcal O}(1/\sqrt{|\nu_k|})$.

\begin{figure}[tbp]
\begin{center}
\scalebox{0.35}{\includegraphics{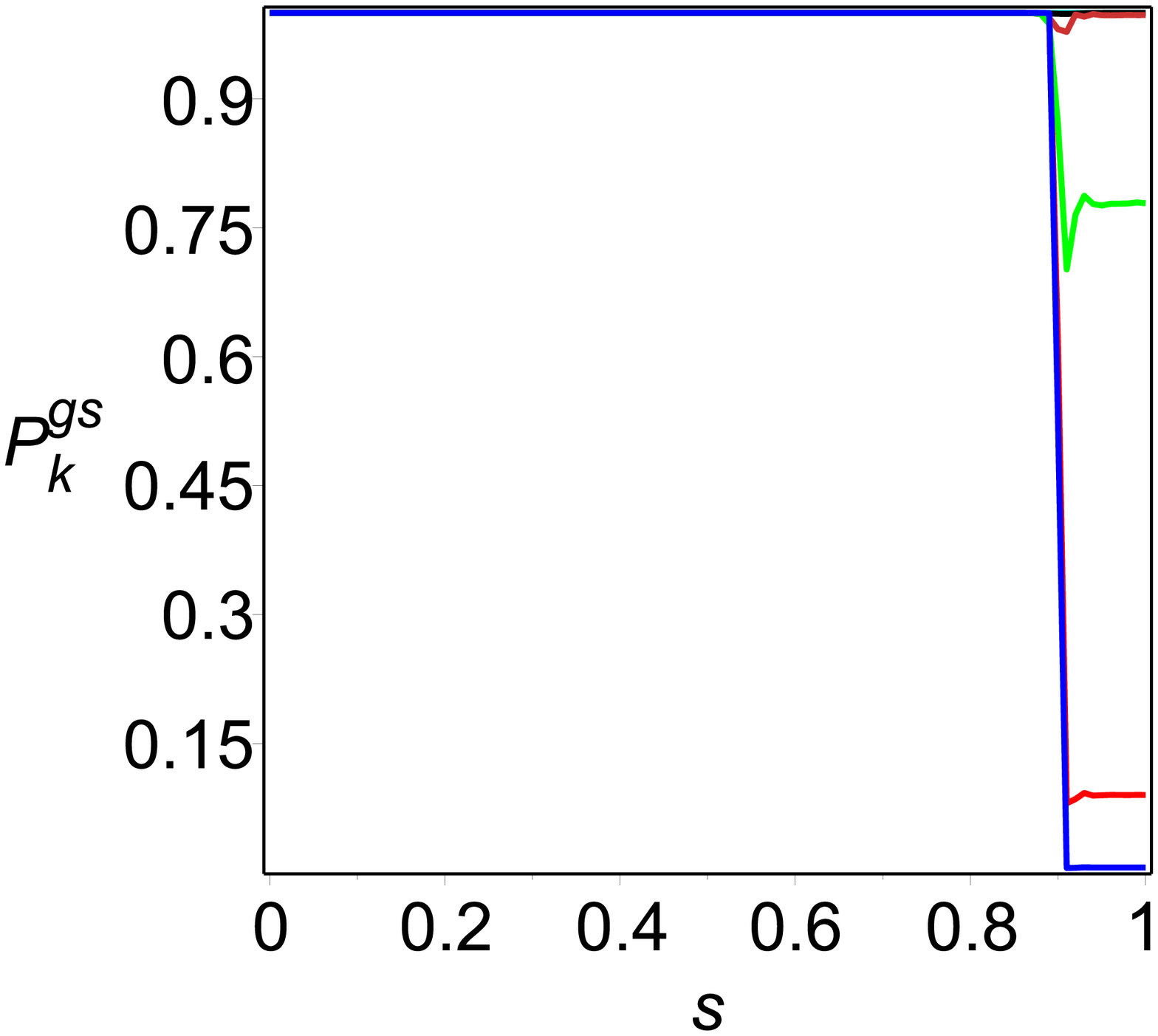}}
\scalebox{0.36}{\includegraphics{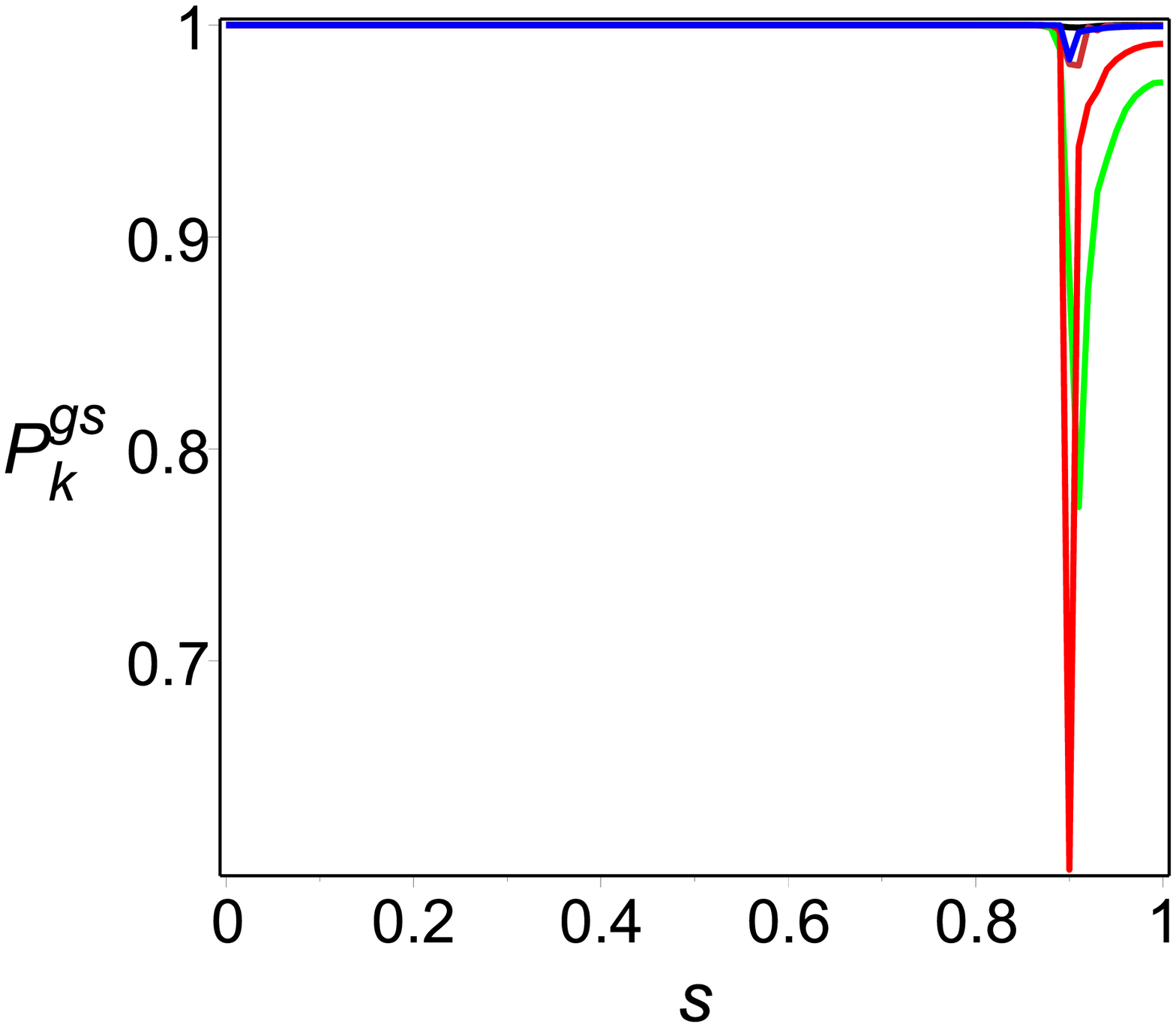}}
\end{center}
\caption{(Color online) Left panel. The probability, $P^{gs}_k$, to remain in the ground state as a function of the scaled time, $s =t/\tau $, for the Hermitian QA ($\delta = 0$, $J=0.5$, $g=10$, $\tau = 10^3$, $N=1024$). From bottom to top: $k=1,4,16,32,64$. Right panel. The probability, $P^{gs}_k$, to remain in the ground state as a function of the scaled time, $s =t/\tau $, for the NQA ($\delta = 0.25$, $J=0.5$, $g=10$, $\tau = 10^3$, $N=1024$). Blue line ($k=1$), red line ($k=4$), green line ($k=16$), orange line ($k=32$), black line ($k= 64$). }
\label{QA1}
\end{figure}

Our theoretical predictions are confirmed by numerical calculations performed for up to $N=1024$ qubits. (See Fig. \ref{QA1}.) One can observe that while short wavelength excitations are essential at the critical point, at the end of evolution their contribution to the transition probability from the ground state to the first excited state is negligible.

Since for each $k$,  the evolution is an independent, the probability of the whole system to remain in the ground state at the end of evolution is the product
\begin{eqnarray} \label{Eq4ra}
P_{gs} = \prod_{k} P^{gs}_k(\tau).
\end{eqnarray}
In the long wavelength approximation one can take into account only $\varphi_k= \pi/N$, and estimate $ P_{gs}$ as
\begin{eqnarray} \label{Eq4a}
P_{gs} \approx \frac{1-  e^{-2\pi\tau/\tau_0}}{1-e^{-2\pi\tau/\tau_0}+  e^{-2\pi\tau/\tau_0- \Re z^2(\tau)}},
\end{eqnarray}
where $\tau_0 = 2gN^2/(\pi^2J)$ and  $\Re z^2(\tau) = 2\delta J\tau/g^2$.

For the Hermitian QA $(\delta =0)$, Eq. (\ref{Eq4a}) yields the Landau-Zener formula \cite{LL,ZC}
\begin{eqnarray}\label{P}
P_{gs} =1-e^{-2\pi\tau/\tau_0}.
\end{eqnarray}
From here it follows that $P_{gs} \approx 1$, if $\tau \geq \tau_0 $. Thus, the computational time for the Hermitian QA should be of order $N^2$.

For the NQA, assuming $\tau \ll \tau_0$, we obtain
\begin{eqnarray}\label{Pq2g}
P_{gs} \approx \frac{1}{1+ \displaystyle\frac{\tau_0}{2\pi\tau}\,  e^{- 2J\delta\tau/g^2}}.
\end{eqnarray}
From here, in the limit of $\delta \rightarrow 0$, we obtain
\begin{eqnarray}\label{Pq2_k}
P_{gs} \rightarrow \frac{1}{1+ \displaystyle\frac{\tau_0}{2\pi\tau}} \ll 1.
\end{eqnarray}
The obtained result is expected, as in this case the time of Hermitian annealing, $\tau$, is small with respect to the characteristic time, $\tau_0$: $\tau\ll \tau_0$.

Next, assuming
\begin{eqnarray}\label{AT1}
\frac {2J \delta\tau}{g^2} - \ln \frac{\tau_0}{2\pi\tau} \gg 1.
\end{eqnarray}
we obtain
\begin{eqnarray}\label{Pq2_g}
P_{gs} \approx 1- \frac{\tau_0}{2\pi\tau}\,  e^{-2J \delta\tau/g^2}.
\end{eqnarray}
As one can see, $P_{gs} \approx 1$, if the conditions of Eq. (\ref{AT1}) are satisfied.
From (\ref{AT1}) we obtain the following rough estimate of the computational time for NQA: $\tau \gtrsim (g^2/2J\delta)\ln N$.

\begin{figure}[tbp]
\begin{center}
\scalebox{0.325}{\includegraphics{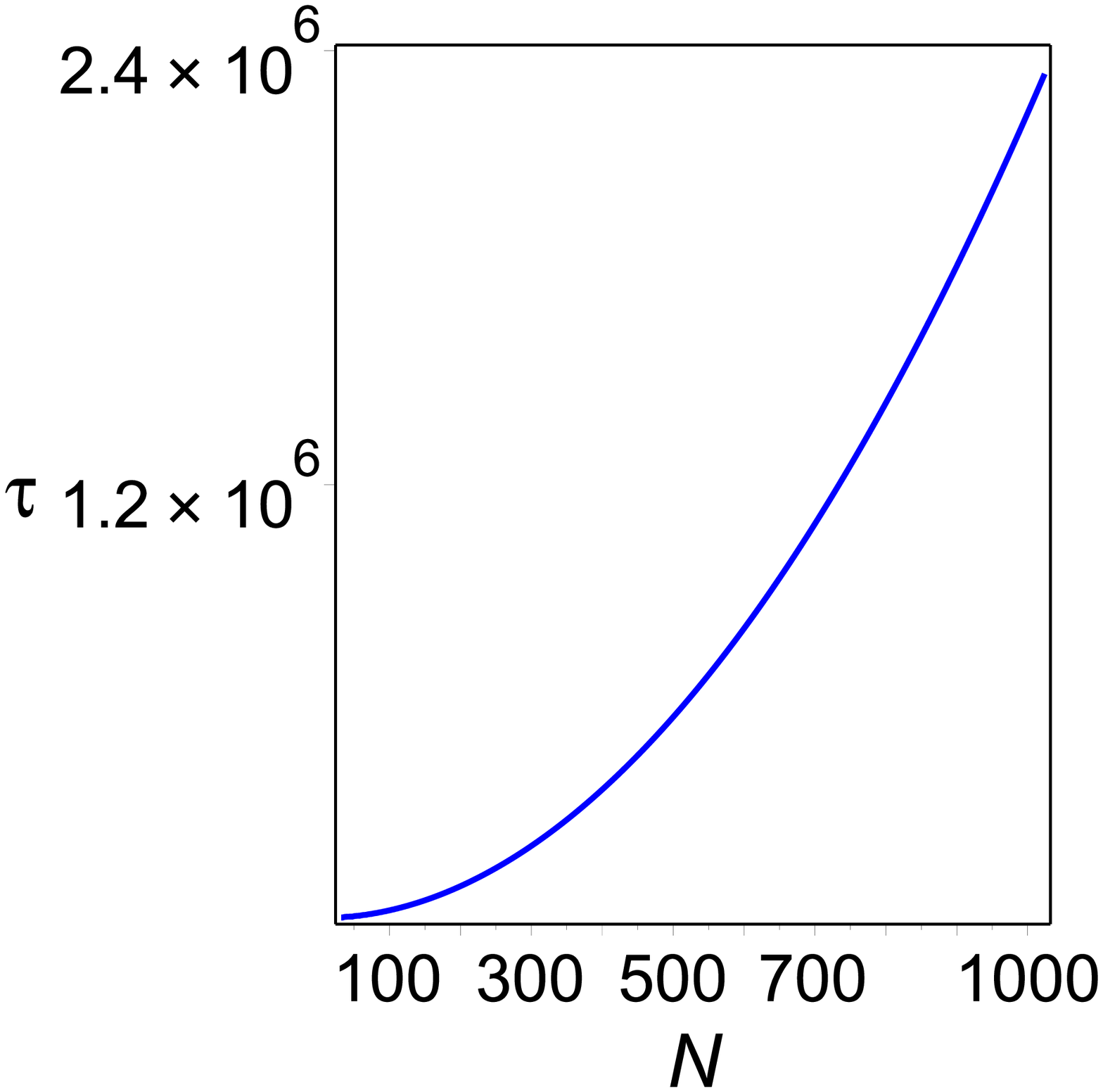}}
\scalebox{0.425}{\includegraphics{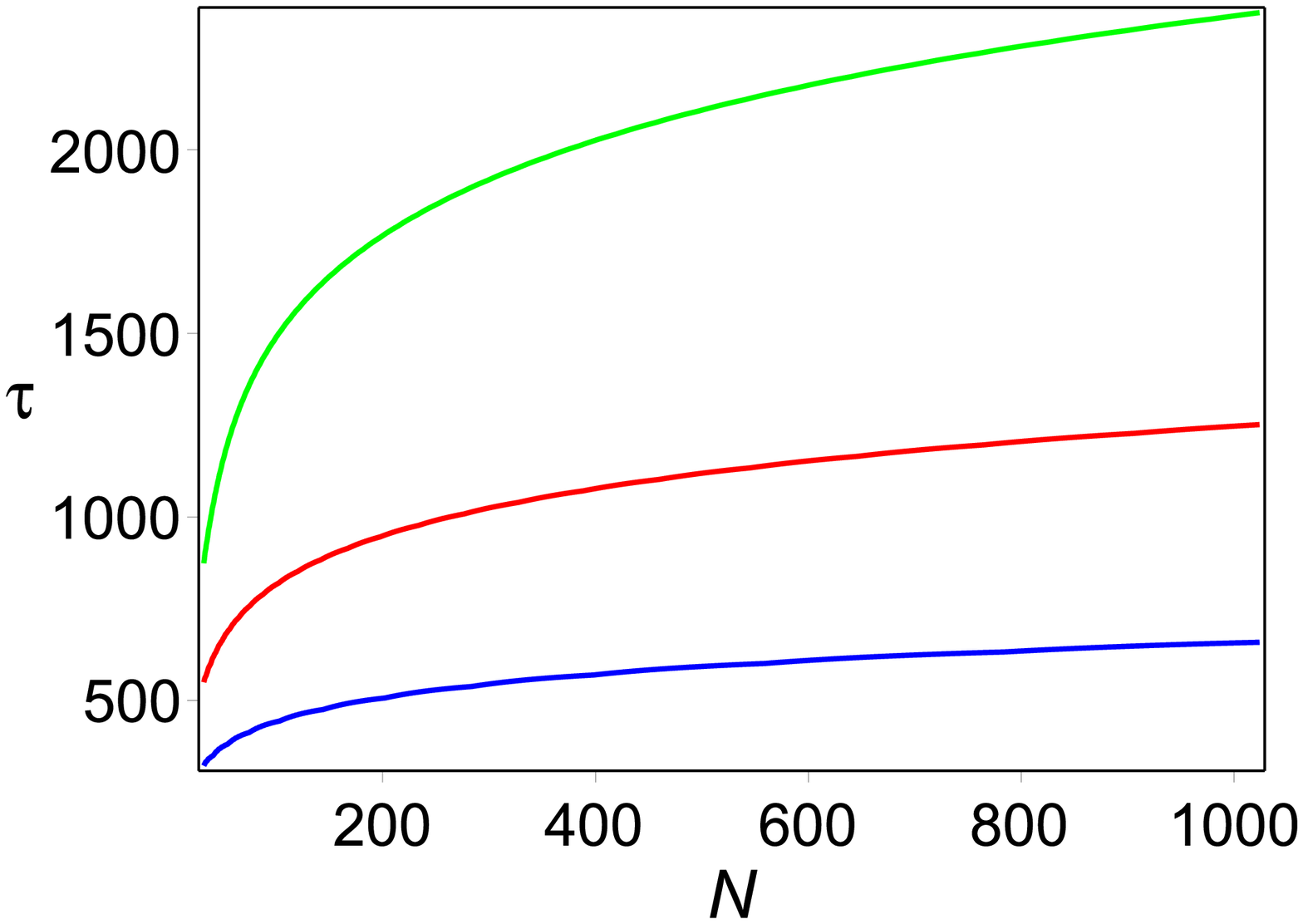}}
\end{center}
\caption{(Color online) Annealing time, $\tau$, as function of $N$. Left panel.  Hermitian QA, $\delta=0$  ($J=0.5$, $g=10$). Right panel. Non-Hermitian QA, from top to bottom: $\delta=0.25,0.5,1 $.}
\label{tau}
\end{figure}

In Fig. \ref{tau}, the annealing time, $\tau$, as a function of the number of spins is shown for the Hermitian QA (left panel) and for the NQA (right panel). The comparison shows that for large number of spins ($N\sim 1000$) and for $\delta\gtrsim 0.25$, the annealing time of the NQA is $\approx 10^{3}$ times smaller than the time for the Hermitian QA.

These results demonstrate that even moderate dissipation increases the transition probability. The characteristic time of non-Hermitian annealing, even for small but finite  , is determined not only by the number of spins,   (as in the Hermitian annealing), but mainly by the dissipation rate,  . The non-Hermitian quantum annealing has a complexity of order  , which is significantly better than the corresponding complexity for the quantum Hermitian adiabatic algorithm ($\sim N^2$).

\section{Correlation functions and defects formation }

During QA the system does not remain in the ground state at all times. At the critical point, the quantum system becomes excited, and its final state is determined by the number of defects. In the case of the antiferromagnetic Ising chain, the system ends in the state such as
\begin{eqnarray}\label{Eq5d}
\large\boldsymbol |\dots \uparrow\downarrow\uparrow\downarrow\downarrow\uparrow\downarrow\uparrow\downarrow\uparrow
\downarrow\uparrow\downarrow\uparrow\uparrow \downarrow\uparrow\downarrow\uparrow\downarrow\uparrow\downarrow\uparrow\downarrow \downarrow\uparrow\downarrow \uparrow \downarrow \uparrow\downarrow\uparrow\downarrow\uparrow\downarrow, \dots \rangle
\end{eqnarray}
with neighboring spins polarized on opposite directions along the $z$-axis and separated by walls (defects) in which the polarization of spins has the same orientation.

The antiferromagnetic spin-spin correlation functions provide information about long-range order and defects formation. We restrict ourselves consider of the antiferromagnetic correlation function only in the final state,
\begin{eqnarray}\label{chiC1}
 \chi(p)=\frac{1}{N}\sum^N_{n=1}({\langle \sigma_n^z \sigma_{n+p}^z\rangle}- {\langle \sigma_n^z \rangle \langle \sigma_{n+p}^z\rangle}),
\end{eqnarray}
defined at $\tilde g(\tau) =0$.

Using the Jordan-Wigner transformation, the correlation function can be recast as follows:
\begin{eqnarray}\label{chiC1a}
 \chi(p)=\frac{1}{N}\sum^N_{n=1}\langle B_n A_{n+1}B_{n+2} A_{n+3}\dots  B_{n+p-1} A_{n+p}\rangle,
\end{eqnarray}
where $A_m =c^\dagger_m+ c_m $ and $B_m =c^\dagger_m - c_m $.

Next, using Wick's theorem, we can express the average in Eq. (\ref{chiC1a}) in terms of contractions of pairs: $\langle B_m  A_n\rangle$,  $\langle B_m B_j\rangle$ and $\langle  A_m  A_n\rangle$  \cite{LSM}. When the pairing  $\langle B_m B_j\rangle =0$ and  $\langle  A_m A_n\rangle =0$ for $m\neq n$, the correlation function can be written as a determinant of the Toeplitz matrix,
$\chi_\tau(p)= \det G$,
where
\begin{eqnarray}\label{G2}
G=\left(
  \begin{array}{cccc}
     G_{0} & G_{-1} & \dots & G_{-p+1} \\
  G_{1} & G_{0} & \dots & G_{-p+2} \\
  \vdots & \vdots & \vdots & \vdots \\
  G_{p-1} & G_{p-2} & \dots & G_{0}
  \end{array}
\right)
\end{eqnarray}

Its components are defined as follows:
\begin{eqnarray}\label{Cor1r}
 G_{p}(\tau)= \frac{1}{N}\sum_{k} \frac{|v_k|^2 -|u_k|^2 -i(u_k v^\ast_k + u^\ast_k v_k )}{|v_k|^2 +|u_k|^2}\,e^{i(m-n)\varphi_k}.
\end{eqnarray}
In the thermodynamic limit, $N \rightarrow \infty$, we obtain
\begin{eqnarray}\label{Cor1}
 G_{k}= \frac{1}{2\pi }\int^{\pi}_{-\pi} \frac{(|v |^2 -|u |^2 -i(u  v^\ast  + u^\ast  v  ))e^{i k\varphi}d\varphi}{|v |^2 +|u |^2}.
\end{eqnarray}

Computation of the matrix elements of the Toeplitz matrix for the ground state of the Hamiltonian, ${\mathcal H}_0 $, yields the following result \cite{LSM,MBM1,BMBM,ZLSD}:
\begin{eqnarray}\label{G1}
G_{k} =\left\{ \begin{array}{cc}
             -1, & k= 0 \\
             0, & k= \pm 1,\pm 2,  \dots
           \end{array}
           \right.
          \end{eqnarray}
Using this result, we obtain the correlation function of the antiferromagnetic chain as $\chi(p) = (-1)^p$.

For the time-dependent problem, the correlation function is not defined by a determinant, since in the general case
\begin{eqnarray}\label{Eq4}
\langle  A_m  A_n\rangle = \delta_{mn} + 2i \Im \beta_{mn}, \\
 \langle B_m B_n\rangle =- \delta_{mn} + 2i \Im \beta_{mn},
\end{eqnarray}
where
\begin{eqnarray}\label{Eq5}
  \beta_{mn} = \frac{1}{2\pi i}\int^{\pi}_{-\pi}\frac{u v^\ast e^{i\varphi(m-n)}d\varphi}{|u|^2 +| v|^2}.
\end{eqnarray}

\begin{figure}[tbp]
\begin{center}
\scalebox{0.65}{\includegraphics{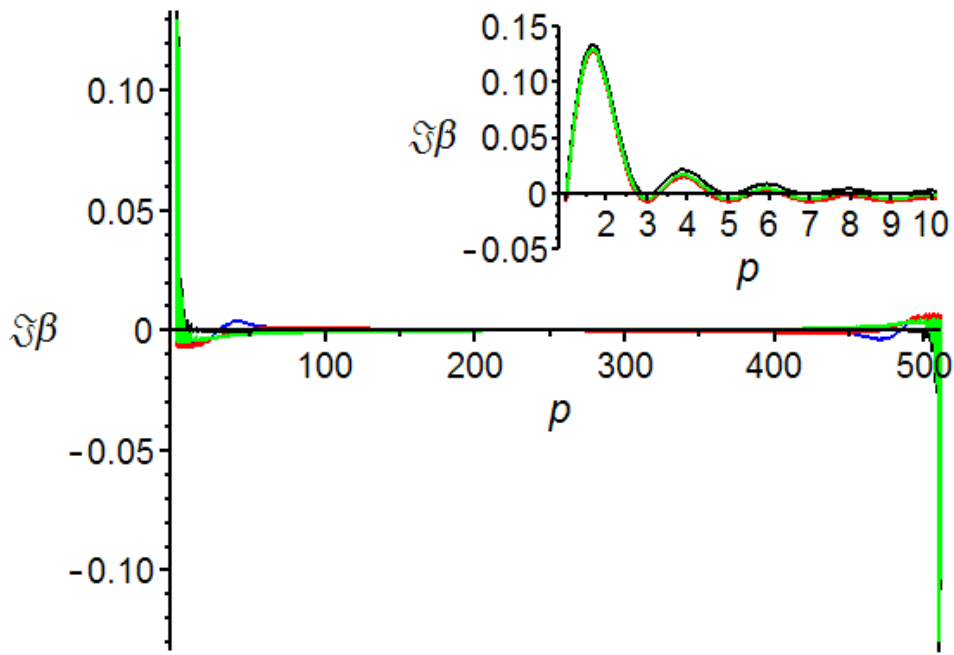}}
\end{center}
\caption{(Color online) Dependence of $ \Im \beta$ on $p=m-n$,  $\delta= 0$ (blue), $\delta= 0.25$ (red), $\delta= 0.5$ (green), $\delta= 1$ (black)  ($\tau=1000$, $g=10$, $J=0.5$, $N=512$). }
\label{chi_512}
\end{figure}

In Fig. \ref{chi_512}, the results of our numerical calculation for $N=512$ are shown. As can be observed (see inset) for the correlation distance, $p \gg 1$, we can neglect the contribution of all $\Im \beta_{mn}$. This is valid for  both Hermitian QA and for NQA, as well. Thus, the long-range correlation function can be calculated using the Toeplitz determinant with
\begin{eqnarray}\label{Cor1ar}
 G_{k}= \frac{1}{2\pi }\int^{\pi}_{-\pi} \frac{(|v |^2 -|u |^2)e^{i k\varphi}d\varphi}{|v |^2 +|u |^2}.
\end{eqnarray}

The asymptotics of the Toeplitz determinant can be obtained by applying the Szeg\"o limit theorem. We use the results of Ref. \cite{CRLS} to obtain for the Hermitian QA ($\delta =0$) the following asymptotic formula for the correlation function:
\begin{eqnarray}\label{Eq5c}
 \chi(p) \sim (-1)^p \exp\bigg(-0.174 \frac{p}{\hat \xi} \bigg) \cos \bigg(\sqrt{\frac{\ln 2}{2\pi}} \frac{p}{\hat \xi}+\varphi_0 \bigg),
\end{eqnarray}
where the phase, $\varphi_0$, depends on the parameter, $\hat \xi=\sqrt{J\tau/2g}$, (for details see Ref. \cite{CRLS}).

\begin{figure}[tbp]
\begin{center}
\scalebox{0.55}{\includegraphics{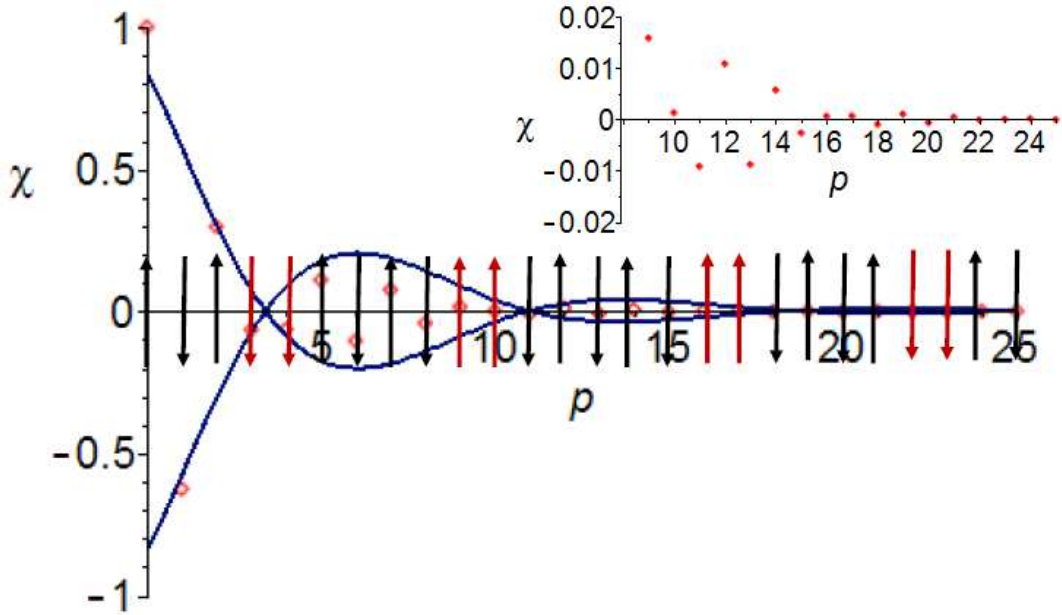}}
\end{center}
\caption{(Color online) Spin antiferromagnetic correlation function and schematic position of spins. Spins, being polarized on opposite directions inside of domain, are separated by walls (red arrows) where the polarization of spins has the same orientation. The correlation length and domains size are determined by the KZ parameter, $\hat \xi$. Red diamonds: correlation function calculated for $N=512$ spins ($\tau =25$, $\delta =0$). Solid line: envelope in Eq. (\ref{Eq5c}).}
\label{spin}
\end{figure}

Asymptotically, the correlation function exhibits decaying oscillations. This agrees with the results obtained in Ref. \cite{CLDJ} for the ferromagnetic Ising chain, where the asymptotic behavior of the correlation function is defined by Eq. (\ref{Eq5c}), without the factor, $(-1)^p$. For the ferromagnetic Ising chain the oscillatory behavior, corresponds to alternating magnetization signs in neighboring ordered domains, separated by kinks \cite{CRLS,CLDJ}. For the antiferromagnetic Ising chain the neighboring spins, being polarized on opposite directions, are separated by walls (defects) where the polarization of spins has the same orientation. In both cases, the Kibble-Zurek (KZ) correlation length \cite{CLDJ}, $\hat \xi$, determines the characteristic domain size as $L= \pi \hat \xi\sqrt{2\pi/\ln 2}$. In Fig. \ref{spin}, the results of our numerical simulations are depicted for $N= 512$ spins and $\tau =25$.
The theoretical prediction of the domain size, $L \approx 7.5$, agrees with the results of the numerical calculation.

\begin{figure}[tbp]
\begin{center}
\scalebox{0.55}{\includegraphics{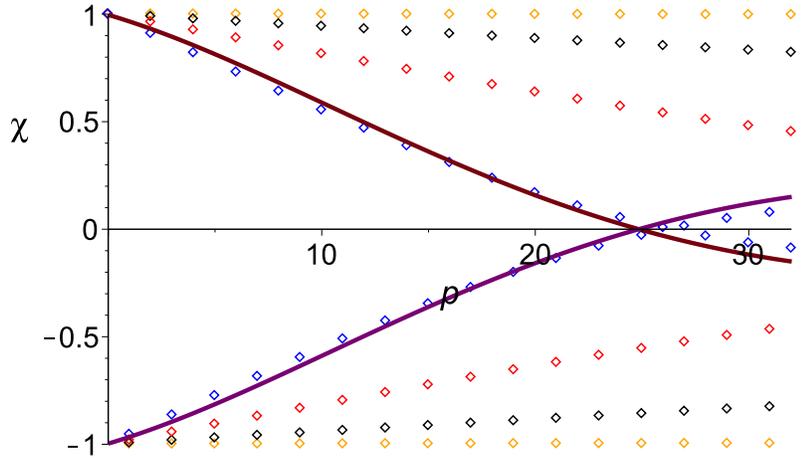}}
\end{center}
\caption{(Color online) Long range behavior of the correlation function at the end of annealing. From bottom to top: $\delta=1,0.5,0.25,0$ ($\tau=1000$, $g=10$, $J=0.5$, $N=512$). Solid lines present the asymptotics given by Eq. (\ref{Eq5c}). }
\label{chi_512a}
\end{figure}

The decaying oscillatory behavior of the correlation functions is confirmed by our numerical calculations presented in Figs. \ref{spin} and \ref{chi_512a}. One can observe that for NQA the long-range behavior of the correlation functions is much better then for the corresponding Hermitian QA, and the characteristic domain size is much longer for NQA than $L$, even for modest values of $\delta$.

To evaluate the efficiency of the QA  one should calculate the number of defects. The computational time of the QA is the time required to achieve the number of defects below some acceptable value. The number of defects is equal to the number of quasiparticles
excited at $\tilde g=0$ (final state) \cite{DJ}. It is given by
${\mathcal N} =  \langle \psi_\tau |\hat{\mathcal N} | \psi_\tau \rangle $,
where
\begin{eqnarray}
 \hat{\mathcal N} =\sum_{k} b_k^\dagger b_k= \frac{1}{2}\sum^N_{n=1}\big ( 1 +\sigma^z_n \sigma^z_{n+1}\big ).
\end{eqnarray}

In the adiabatic regime we can approximate the average number of defects as,
\begin{eqnarray}
\bar {\mathcal N}=1 -P_{k}(\tau),
\end{eqnarray}
with the lowest $\varphi_k = \pi/N$ \cite{CLDJ}. Using Eq. (\ref{Eq4a}), we obtain
\begin{eqnarray} \label{Eq4f}
\bar {\mathcal N} = \frac{ e^{-2\pi\tau/\tau_0- \Re z^2(\tau)}}{1-e^{-2\pi\tau/\tau_0}+  e^{-2\pi\tau/\tau_0- \Re z^2(\tau)}}.
\end{eqnarray}
When $\tau \gg \tau_0$ we find
\begin{eqnarray} \label{Eq4fr}
\bar {\mathcal N} \approx \exp\bigg (-\frac{2 \pi \tau}{\tau_0}\bigg ).
\end{eqnarray}
So, for large annealing times, the NQA does not have any advantage in comparison with the Hermitian QA.

In the opposite limit,  $\tau \ll \tau_0$, and under the condition
\begin{eqnarray}\label{AT1a}
\frac {2J \delta\tau}{g^2} - \ln \frac{\tau_0}{2\pi\tau} \gg 1,
\end{eqnarray}
the average number of defects is approximated by
\begin{eqnarray} \label{Eq4fh}
\bar {\mathcal N} \approx  \frac{\tau_0}{2\pi\tau}\,  e^{- 2J\delta\tau/g^2}.
\end{eqnarray}
For the parameters chosen in our numerical simulations: $N=512$, $J=0.5$, $g=10$, and for $\tau = 10^3$ we obtain $\bar {\mathcal N}\approx 40 e^{-10\delta}$. Thus, while for the Hermitian QA, the number of defects, separating the domains with antiferromagnetic order, is $\bar {\mathcal N}\approx 40$, for the NQA with the same protocol, the number of defects decreases as $e^{-10\delta}$. For instance, for $\delta=0.25$ we obtain $\bar {\mathcal N} \approx 3$.

Using Eq. (\ref{chiC1}), we can calculate the density of defects in terms of the correlation function as follows:
\begin{eqnarray}\label{TL1}
n= \lim_{N\rightarrow \infty} \frac{\mathcal N}{N}  = \frac{1}{2}(1+\chi(1)).
\end{eqnarray}
As shown in the previous section, during the slow evolution only long wavelength modes can be excited. So, one can use the Gaussian distribution by replacing $\sin \varphi \approx \varphi $ and $\cos \varphi_k \approx 1$. In the limit $\sqrt{2J\tau/g} \gg 1$, we can employ Eqs. (\ref{Pq2k}) - (\ref{Eq3b}) to calculate the density of defects,
\begin{eqnarray}\label{QP1}
n=  \frac{1}{\pi}\int^{\pi}_{0}  \frac{ e^{-2\pi\Re\nu- \Re z^2}d \varphi }{1-e^{-2\pi\Re\nu}+  e^{-2\pi\Re\nu- \Re z^2}},
\end{eqnarray}
where $\varphi_k \rightarrow \varphi $ as $N\rightarrow \infty$. Performing the integration with $\Re\nu = J\tau \varphi^2 /2g$  and $ \Re z^2= 2\delta J\tau/g^2$, we obtain
\begin{eqnarray}\label{Eq10h}
n =  n_0 e^{-2\delta\tau J/g^2}\Phi\Big( 1-e^{-2\delta\tau J/g^2},\frac{1}{2},1\Big),
\end{eqnarray}
where
\begin{eqnarray}\label{n1}
 n_0 = \frac{1}{2\pi}\sqrt{\frac{g}{J\tau}}
\end{eqnarray}
denotes the density of defects for the Hermitian LZ problem \cite{DJ}, and  $\Phi(x,a,c)$ is the Lerch transcendent \cite{EMOT1}.
\begin{figure}[tbp]
\begin{center}
\scalebox{0.5}{\includegraphics{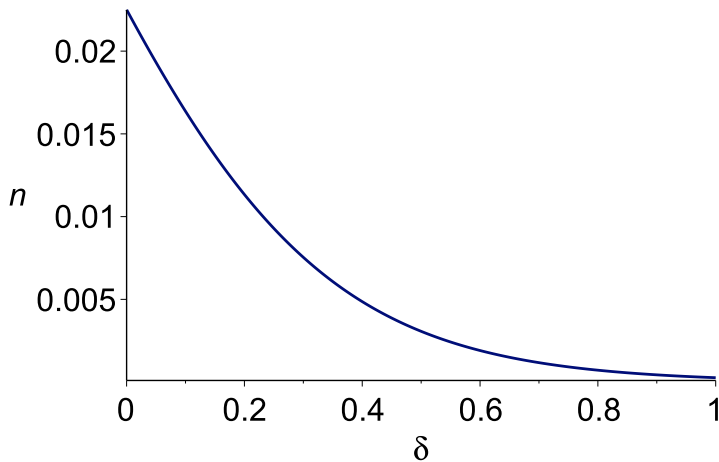}}
\end{center}
\caption{(Color online) Density of defects for NQA as a function of the dissipation parameter, $\delta$ ($\tau = 10^3$, $J=0.5$, $g=10$).}
\label{P2ka}
\end{figure}

In Fig. \ref{P2ka} the density of defects as a function of the decay parameter, $\delta$, is shown. As one can see, even moderate dissipation greatly decreases the number of defects in the system.

\section{Conclusion }

Previously, research on quantum annealing could be conventionally divided by two parts: (i) combining classical computers with quantum algorithms and by (ii) building  real quantum computers. Many schemes and approaches  of quantum annealing algorithms have been proposed \cite{KN,FGGLL,SSMO,DC,SMTC,SST,SNS,AM,OMNH,MO,ON} (see also references therein). The main objective of these publications is to significantly decrease the time of annealing. The very popular test models are the one-dimensional Ising spin chains, ferromagnetic and antiferromagnetic, which  are also useful for practical purposes. In this case, the quantum annealing algorithms are used to find the ground state of the chain.

The approach presented in this paper is related to item (i) above, and was applied for the one-dimensional antiferromagnetic spin chain. We have chosen an auxiliary Hamiltonian in such a way that the total Hamiltonian is non-Hermitian. This allows us to shift the minimal gap in the energy spectrum in the complex plane, and to significantly reduce the time required to find the ground state. Our approach leads to an annealing time proportional to: $ \ln N$, where $N$ is the number of spins. This is much less than the time of Hermitian annealing ($\sim N^2$) for the same problem. We also demonstrated the behavior of quantum correlation functions and their relation to the spin defects generated during quantum annealing.

It is interesting to mention here, that similar to the ferromagnetic chain, the main contribution to the destruction of the final ground state in the process of quantum annealing is produced by the long-wavelength modes, as expected. This is in spite of the fact that in the antiferromagnetic chain the ground state represents a rapidly-oscillating spatial structure with short wavelengths (opposite to the ferromagnetic chain).

The dissipative term which we use corresponds to tunneling of the system to its own continuum, as usually happens when one applies the Feshbach projection method to intrinsic states in nuclear physics and quantum optics. In our case, the intrinsic states are the states of the quantum computer (register). So, the probability of our quantum computer to survive during the NQA can be small. That is why we use a ratio of two probabilities -- the ratio of the probability for the system to remain in the ground state to the probability of survival of the quantum computer. This relative probability (we call it ``intrinsic'' probability) is well-defined, and remains finite during the NQA. So, our approach cannot be directly used in the experiments on QA, but rather as a combination of classical computer and protocols for NQA to significantly decrease the time of annealing. Also, the dissipative term which we use in this paper is rather artificial in the sense that it has no direct relation to physical dissipative mechanisms. At the same time, we note that our dissipation term corresponds, in principal, to tunneling effects in the superconducting phase qubits if tunneling is realized mostly from the lowest energy levels.

Our hope is that the observed reduction of the annealing time can serve as a guide to finding solutions of the NP-complete problems by using a combination of classical computers and NQA algorithms. We also would like to mention that many important problems still remain to be considered. One of them is the application of this dissipative approach to more complicated Ising models with frustrated interactions, in which the ground state can be strongly inhomogeneous. Another approach (``Ising machine") was proposed recently in \cite{SU,TUY} in which both energy dissipation and pumping effects were utilized for mapping of Ising models onto laser systems. In this case, the system can work in the regime of a limit cycle which could be advantageous for performing quantum algorithms. Both of these approaches to applying NQA algorithms to Ising models are now in progress.

\ack
\addcontentsline{toc}{section}{Acknowledgments}

The work by G.P.B. and A.R.B. was carried out under the auspices of the National Nuclear Security Administration of the U.S. Department of Energy at Los Alamos National Laboratory under Contract No. DE-AC52-06NA25396. A.I.N. acknowledges the support from the CONACyT, Grant No. 15439. J.C.B.Z. acknowledges the support from the CONACyT, Grant No. 171014.

\appendix

\section{}

\section*{Exact solution of the Non-Hermitian Landau-Zener problem}

The non-Hermitian Hamiltonian, $ \mathcal {H}_k(t)$, projected on the two-dimensional subspace spanned by $|k_1\rangle= {\scriptsize \left(
                      \begin{array}{c}
                        1 \\
                        0 \\
                      \end{array}
                    \right)}$ and $|k_0\rangle ={\scriptsize \left(
                      \begin{array}{c}
                        0 \\
                        1 \\
                      \end{array}
                    \right)}$, takes the form
\begin{eqnarray}\label{AH1}
 \mathcal {H}_k(t)  = -\varepsilon_0(t) {1\hspace{-.125cm}1} -J \left(
\begin{array}{cc}
              \tilde g(t)- \cos \varphi_k & \sin \varphi_k \\
             \sin \varphi_k & -\tilde g(t)+ \cos \varphi_k   \\
            \end{array}
          \right),
\end{eqnarray}
where $\varepsilon_0(t)= J\cos \varphi_k+ iJ\delta(t)$ and $\tilde g(t) = g(t) - i\delta(t)$.
We assume a linear dependence of the function, $\tilde g(t)$, on time:
\begin{eqnarray}
\tilde g(t) = \left\{
\begin{array}{l}
\gamma (\tau-t), \quad 0 \leq t \leq \tau \\
0 , \quad t > \tau
\end{array}
\right.,
\end{eqnarray}
where, $\gamma=(g-i\delta)/\tau$, and $g$, $\delta$ are real parameters.

The general wave functions,  $|\psi_k\rangle$ and  $\langle\tilde \psi_k |$, satisfy the Schr\"odinger equation and its adjoint equation
 \begin{eqnarray}\label{AEqh1}
i\frac{\partial }{\partial t}|\psi_k\rangle&= \mathcal {H}_k(t)|\psi_k\rangle , \\
-i\frac{\partial }{\partial t}\langle\tilde \psi_k |& =
\langle\tilde \psi_k |\mathcal {H}_k(t)\label{NS_2}.
\end{eqnarray}
Presenting $|\psi_k(t)\rangle$ as a linear superposition
\begin{eqnarray}\label{AS1}
|\psi_k(t)\rangle = (u_k(t)|k_0\rangle + v_k(t)|k_1\rangle) e^{i\int \varepsilon_0(t)dt},
\end{eqnarray}
and inserting (\ref{AS1}) into Eq. (\ref{AEqh1}), we obtain
\begin{eqnarray}\label{AS2a}
i\dot  u_k &= J\big((\tilde  g - \cos \varphi_k)\,u_k - \sin \varphi_k\, v_k\big), \\
i\dot  v_k &= -J\big(\sin \varphi_k\, u_k +(\tilde  g - \cos \varphi_k)\,v_k \big ).
\label{AS2b}
\end{eqnarray}

Let $z_k(t) = e^{i\pi/4}\sqrt{2J/\gamma}\big(\gamma(\tau -t)-\cos \varphi_k\big)$ be a new variable. Then, for new functions, $u_k(t)= U_k(z_k)$ and $v_k(t) = V_k(z_k)$, we obtain
\begin{eqnarray}\label{AS3a}
\frac{d}{dz_k}U_k  =  \frac{z_k}{2}U_k -\sqrt{i\nu_k}V_k, \\
\frac{d}{dz_k}V_k  =  -\frac{z_k}{2}V_k -\sqrt{i\nu_k}U_k,
\label{AS3b}
\end{eqnarray}
where $\nu_k= (J/2\gamma)\sin^2 \varphi_k$, and the complex `time' $z_k$ runs from $z_k(0) = e^{i\pi/4}\sqrt{2J/\gamma}\big(\gamma \tau-\cos \varphi_k\big)$ to $z_k(\tau) = -e^{i\pi/4}\sqrt{2J/\gamma}\cos \varphi_k$.

From Eqs. (\ref{AS3a}), (\ref{AS3b}), we obtain the second order Weber equations
\begin{eqnarray}
\frac{d^2}{dz^2_k}U_k  - \Big( \frac{1}{2} + \frac{z_k^2}{4}+i\nu_k\Big)U_k =0 ,\\
\frac{d^2}{dz^2_k}V_k + \Big(\frac{1}{2} -\frac{z_k^2}{4} -i\nu_k\Big)V_k =0 .
\end{eqnarray}
Solution of the Weber's equations is given by the parabolic cylinder functions, $D_{-i\nu_k}(\pm z)$ and $D_{i\nu_k-1}(\pm i z)$.

We obtain the solutions of Eqs. (\ref{AS3a}, \ref{AS3b}) in the form
 \begin{eqnarray}\label{AEq4}
& U_{k}(z_k)= B_k D_{i\nu_k}(iz_k) +{\sqrt{i\nu_k }}  A_k D_{-i\nu_k-1}(z_k)  ,\\
& V_{k}(z_k)=   A_k D_{-i\nu_k}(z_k) -i\sqrt{i\nu_k } B_k D_{i\nu_k-1}(iz_k))  ,
\label{AEq4a}
 \end{eqnarray}
where the constants, $A_k$ and $B_k$, are determined from the initial conditions.

\end{document}